\def\zem{$z_{\rm em}$}
\def\zabs{$z_{\rm abs}$}
\def\la{$_<\atop{^\sim}$}
\def\ga{$_>\atop{^\sim}$}
\def\kms{km~s$^{-1}$}

\def\cm2{cm$^{-2}$}
\def\loghi{\mbox{$\log N_{\rm HI}$}}
\def\nhi{\mbox{$N_{\rm HI}$}}
\def\hi{H~{\sc i}}

\def\e{et~al.}

\def\c2{C~{\sc ii}}
\def\c4{C~{\sc iv}}
\def\fe2{Fe~{\sc ii}}
\def\fe3{Fe~{\sc iii}}
\def\mg1{Mg~{\sc i}}
\def\mg2{Mg~{\sc ii}}
\def\n1{N~{\sc i}}
\def\h1{H~{\sc i}}

\def\approxlt{\mathrel{\spose{\lower 3pt\hbox{$\sim$}}
        \raise 2.0pt\hbox{$<$}}}
\def\approxgt{\mathrel{\spose{\lower 3pt\hbox{$\sim$}}
        \raise 2.0pt\hbox{$>$}}}

\def\ion#1#2{{\rm #1}\,{\sc #2}} 
\documentstyle[psfig,rotating]{mn}

\input{psfig.sty}
\newif\ifAMStwofonts

\ifoldfss
  \ifCUPmtlplainloaded \else
    \NewTextAlphabet{textbfit} {cmbxti10} {}
    \NewTextAlphabet{textbfss} {cmssbx10} {}
    \NewMathAlphabet{mathbfit} {cmbxti10} {} 
    \NewMathAlphabet{mathbfss} {cmssbx10} {} 
  \fi
  \ifAMStwofonts
    \ifCUPmtlplainloaded \else
      \NewSymbolFont{upmath} {eurm10}
      \NewSymbolFont{AMSa} {msam10}
      \NewMathSymbol{\upi}     {0}{upmath}{19}
      \NewMathSymbol{\umu}     {0}{upmath}{16}
      \NewMathSymbol{\upartial}{0}{upmath}{40}
      \NewMathSymbol{\leqslant}{3}{AMSa}{36}
      \NewMathSymbol{\geqslant}{3}{AMSa}{3E}

      \let\geq=\geqslant 
    \fi
  \fi
\fi 

\ifnfssone
  \newmathalphabet{\mathit}
  \addtoversion{normal}{\mathit}{cmr}{m}{it}
  \addtoversion{bold}{\mathit}{cmr}{bx}{it}
  \newmathalphabet{\mathbfit} 
  \addtoversion{normal}{\mathbfit}{cmr}{bx}{it}
  \addtoversion{bold}{\mathbfit}{cmr}{bx}{it}
  \newmathalphabet{\mathbfss} 
  \addtoversion{normal}{\mathbfss}{cmss}{bx}{n}
  \addtoversion{bold}{\mathbfss}{cmss}{bx}{n}
  \ifAMStwofonts
    \ifCUPmtlplainloaded \else
      \UseAMStwoboldmath
      \makeatletter
      \new@mathgroup\upmath@group
      \define@mathgroup\mv@normal\upmath@group{eur}{m}{n}
      \define@mathgroup\mv@bold\upmath@group{eur}{b}{n}
      \edef\UPM{\hexnumber\upmath@group}
      \new@mathgroup\amsa@group
      \define@mathgroup\mv@normal\amsa@group{msa}{m}{n}
      \define@mathgroup\mv@bold\amsa@group{msa}{m}{n}
      \edef\AMSa{\hexnumber\amsa@group}
      \makeatother
      \mathchardef\upi="0\UPM19
      \mathchardef\umu="0\UPM16
      \mathchardef\upartial="0\UPM40
      \mathchardef\leqslant="3\AMSa36
      \mathchardef\geqslant="3\AMSa3E

      \let\geq=\geqslant 
    \fi
  \fi
\fi 

\ifnfsstwo
  \DeclareMathAlphabet{\mathbfit}{OT1}{cmr}{bx}{it}
  \SetMathAlphabet\mathbfit{bold}{OT1}{cmr}{bx}{it}
  \DeclareMathAlphabet{\mathbfss}{OT1}{cmss}{bx}{n}
  \SetMathAlphabet\mathbfss{bold}{OT1}{cmss}{bx}{n}
  \ifAMStwofonts
    \ifCUPmtlplainloaded \else
      \DeclareSymbolFont{UPM}{U}{eur}{m}{n}
      \SetSymbolFont{UPM}{bold}{U}{eur}{b}{n}
      \DeclareSymbolFont{AMSa}{U}{msa}{m}{n}
      \DeclareMathSymbol{\upi}{0}{UPM}{"19}
      \DeclareMathSymbol{\umu}{0}{UPM}{"16}
      \DeclareMathSymbol{\upartial}{0}{UPM}{"40}
      \DeclareMathSymbol{\leqslant}{3}{AMSa}{"36}
      \DeclareMathSymbol{\geqslant}{3}{AMSa}{"3E}

      \let\geq=\geqslant 
    \fi
  \fi
\fi 

\ifCUPmtlplainloaded \else
  \ifAMStwofonts \else 
    \def\upi{\pi}
    \def\umu{\mu}
    \def\upartial{\partial}
  \fi
\fi

\title[Metal-rich DLAs/sub-DLAs at $z<1$]{Metal-rich Damped/sub-Damped Lyman-$\alpha$ Quasar Absorbers at $z<1$
 \thanks{Based on observations collected during programme ESO
 74.A-0597 at the European Southern Observatory with UVES on the 8.2 m
 KUEYEN telescope operated at the Paranal Observatory, Chile.} }

\author[C. P\'eroux et al.] {C. P\'eroux$^1$\thanks{e-mail:
        cperoux@eso.org}, J. D. Meiring$^2$, V. P. Kulkarni$^2$,
R. Ferlet$^3$, P. Khare$^4$, J. T. Lauroesch$^5$,
\newauthor
 G. Vladilo$^6$,  \& D. G. York$^7$.\\
$^1$ European Southern Observatory, Garching-bei-M\"unchen, Germany. \\
$^2$ Dept. of Physics and Astronomy, Univ. of South Carolina, Columbia, USA.\\
$^3$ Institut d'Astrophysique de Paris, UMR7095 CNRS,
Universite Pierre \& Marie Curie, France.\\
$^4$ Dept. of Physics, Utkal University, Bhubaneswar, India.\\
$^5$ Dept. of Physics and Astronomy, Northwerstern University, Evanston, USA.\\
$^6$ Osservatorio di Trieste, Trieste, Italy.\\
$^7$ Dept. of Astronomy and Astrophysics, Univ. of Chicago, Chicago, USA.
}
 \date{Accepted 2006 July 25. Received 2006 July 24; in original form 2006 April 04}

\pubyear{2006}

\begin{document}

\maketitle


\begin{abstract}

Damped Lyman-$\alpha$ absorbers (DLAs), seen in absorption against a
background quasar, provide the most detailed probes available of
element abundances in the Universe over $> 90 \%$ of its age. DLAs can
be used to observationally measure the global mean metallicity in the
Universe and its evolution with time. Paradoxically, these
observations are more difficult at lower redshifts, where the absorber
rest-frame UV spectra are cut-off due to the atmospheric absorption.
We present here high-resolution VLT/UVES observations of several
elements contained in three DLAs and one sub-DLA with
$0.6<$\zabs$<0.9$. We detect Mg I, Mg II, Fe II, Zn II, Cr II, Mn II,
Ti II and Ca II. Our observations more than double the high-resolution
sample of [Zn/H] at $z<1$. We also report the discovery of three
metal-rich systems, whereas most previous measurements show low
\nhi-weighted mean metallicity projecting to about 1/6$^{th}$ solar
level at z=0. We derive [Zn/H]=$-$0.11$\pm$0.04 at \zabs=0.725,
[Zn/H]=$-$0.54$\pm$0.20 at \zabs=0.740 and [Zn/H]=$-$0.49$\pm$0.22 at
\zabs=0.652, plus one additional upper limit ([Zn/H]$<-$0.36 at
\zabs=0.842). These measurements confirm the existence of quasar
absorbers with relatively high metallicities based on abundance
estimates free from the effect of dust depletion.  Possible
implications of these results for the metallicity of neutral gas phase
in the past $\approx 8$ Gyr are presented and compared with models.

\end{abstract}

\begin{keywords}
Galaxies: abundances -- intergalactic medium -- quasars: absorption
   lines -- quasars: individual: SDSS J0134$+$0051, SDSS J0256$+$0110,
   SDSS J1107$+$0048, SDSS J2328$+$0022
\end{keywords}

\section{Introduction}

Damped Lyman-$\alpha$ systems (DLAs) seen in absorption in the spectra
of background quasars are selected over all redshifts independent of
the intrinsic luminosities of the underlying galaxies. They have
hydrogen column densities, \loghi\ $\ga$ 20.3 and are major
contributors to the neutral gas in the Universe (P\'eroux \e\
2003). The DLAs offer direct probes of element abundances over $ > 90
\%$ of the age of the Universe. In particular, the element Zn is only
mildly depleted onto dust grains and is known to trace Fe in most
Galactic stars with [Fe/H]$\geq -$3 (e.g. Mishenina et al. 2002;
Cayrel et al. 2004). Zinc is thus an excellent tracer of the total
metallicity in quasar absorbers (e.g., Pettini, et al. 1997; Kulkarni,
Fall, \& Truran 1997; P\'eroux et al., 2002; Khare et al. 2004). Iron,
on the other hand, is often depleted out of gas phase onto dust
grains, and so, is not a good tracer of metallicity in the
interstellar medium of other galaxies. By comparing the gas phase
abundance of refractory elements such as Fe, Mn, Cr, etc. to that of
Zn, one can assess the quantity of interstellar grains in the systems.

Most chemical evolution models (e.g., Malaney \& Chaboyer, 1996; Pei,
Fall, \& Hauser, 1999) predict the global metallicity of the Universe
to rise from nearly zero at high $z$ to nearly solar at $z=0$ although
galaxy to galaxy variation may be expected from the mass-metallicity
relation (Tremonti et al. 2004; Savaglio et al. 2005) as well as from
gradients of abundance with galaxy radius (i.e. Ellison, Kewley \&
Mall\'en-Ornelas, 2005; Chen et al. 2005; Zwaan et al. 2005). The
present-day mass-weighted mean metallicity of local galaxies is also
nearly solar (e.g., Kulkarni, \& Fall, 2002; Fukugita \& Peebles
2004). The \nhi-weighted metallicity of DLAs is expected to follow
this behaviour if they trace an unbiased sample of absorbers selected
only by $N({\rm H I})$. However, studies of the cosmological evolution
of the \hi\ column density-weighted mean metallicity in DLAs shows
surprising results. Contrary to most models of cosmic chemical
evolution, recent observations suggest a relatively mild evolution in
DLA global metallicity with redshift for 0$\la$z$\la$ 4 (Prochaska \e\
2003; Khare, et~al., 2004; Kulkarni, et~al., 2005; Meiring et~al. 2006
and references therein). Treating Zn limits with survival analysis,
Kulkarni et al. (2005) obtained a slope of $-0.18
\pm 0.07$ and an intercept (i.e., projected metallicity at $z = 0$) of
$-0.79 \pm 0.18$ for the metallicity-redshift relation. Similar values
for the slope have been reported from previous Zn samples ($-0.26 \pm
0.10$; Kulkarni \& Fall 2002), and from a larger heterogeneous sample
of Zn, Fe, Si, S, and X-ray absorption measurements ($-0.25 \pm 0.07$;
Prochaska et al. 2003). Overall the \nhi-weighted mean metallicity of
DLAs is less than 1/10$^{th}$ solar at z=2 and about 1/6$^{th}$ solar
at z=1 (Meiring et al. 2006).

These results suggest that the global mean DLA metallicity stays far
below the solar level at all redshifts.  However, the exact shape of
the metallicity-redshift relation is still not accurately
determined. This uncertainty is mainly due to the small number of
measurements available at $z < 1.5$, which corresponds to $\sim 70 \%$
of the age of the Universe. In fact, most of the low-redshift data
available so far are based on low-resolution spectroscopy. Only 2 DLAs
with $z < 1$ have been measured at high resolution, a DLA at z=0.68
(de La Varga \& Reimers 2000) and a DLA at z=0.86 (Pettini et
al. 2000). This is primarily because (a) for $z < 0.6$, Zn II lines
lie in the UV and can only be accessed with HST (which currently has
no high-resolution spectrograph working), and (b) for $0.6 < z < 1.5$,
Zn II lines lie in the blue wavelengths where most spectrographs have
relatively low sensitivities.  Yet, high resolution is important both
for lowering the detection threshold and for resolving blends of Mg I
and Cr II and the two Zn II lines at $\lambda \lambda$ 2026 2062 which
are separated by 50 km/s and 62 km/s respectively. Fortunately, new
low-redshift DLAs are being discovered rapidly. The 100,000 quasar
spectra being obtained in the Sloan Digital Sky Survey (SDSS) combined
with HST observations (and future facilities) are expected to increase
the number of DLAs by an order of magnitude (York et al. 2000; Rao \&
Turnshek 2000; York, et al. 2001; York et al. 2005). It is important
to determine the elemental abundances in these DLAs which probe a
large fraction of the age of the Universe.

\begin{table*}
\begin{center}
\caption{Dates of observations of our programme to study the 
total metallicity (free from the effect of dust bias) of \zabs$<1$
quasar absorbers.
\label{t:JoO}}
\begin{tabular}{cccccccc}
\hline
\hline
Quasar &g mag &$z_{\rm em}$ &$z_{\rm abs}$ &log \nhi$^a$ &Obs Date
&UVES settings (nm)&Exp. Time (sec)\\
\hline
SDSS J0134$+$0051 &18.33 &1.522 &0.842  &$19.93^{+0.10}_{-0.15}$ &19/21 Oct 2004 &346+564 &3$\times$4800\\
SDSS J0256$+$0110 &18.96 &1.348 &0.725  &$20.70^{+0.11}_{-0.22}$ &21/22 Oct 2004 + 11 Nov 2004&390+564 &3$\times$5400 \\
SDSS J1107$+$0048 &17.66 &1.388 &0.740  &$21.00^{+0.02}_{-0.05}$ &11 Jan 2005 + 1 Mar 2005 &390+564 &2$\times$3600\\
SDSS J2328$+$0022 &18.01 &1.309 &0.652  &$20.32^{+0.06}_{-0.07}$ &02 Oct 2004 &390+564 &3$\times$5300 \\       
\hline
\hline
\end{tabular}
\end{center}
\vspace{0.2cm}
\begin{minipage}{140mm}
{\bf $^a$:} \nhi\ measurements are from Rao, Turnshek, \& Nestor (2006) 
for SDSS J0256$+$0110, SDSS J1107$+$0048, and SDSS J2328$+$0022, and from 
Rao (2005, private communication) for SDSS J0134$+$0051.\\
\end{minipage}
\end{table*}

Here, we present VLT/UVES high-resolution spectra for 4 absorbers at
$z < 1$ which allow us to study their velocity structure, to discern
potential saturation effects, and thus to determine the variation of
metallicity and dust content among the various velocity
components. The paper is organised as follows: section 2 refers to the
target selection and data reduction processes. The analysis of every
individual system is detailed in section 3, while section 4 describes
the abundance determination and potential impact on our understanding
of the metallicity evolution of the Universe.

\section{Observational Strategy and Data Reduction}

In order to assess the metallicity of the Universe at z$<$1, we have
started in autumn 2002 a campaign aimed at measuring abundances in
low-$z$ systems. We have considered the low-$z$ DLAs and sub-DLA
systems (i.e. systems with \loghi$>$19.0 as first defined by P\'eroux
et al. 2003) with known \nhi\ discovered with either the Hubble Space
Telescope ($0.1 < z < 0.6$) or Sloan ($0.6 < z < 1.5$) and followed up
with the Multiple Mirror Telescope (MMT), e.g., Khare et al. 2004;
Kulkarni et al. 2005; Meiring et al. 2006. These systems were chosen
in the first place because their \nhi\ is known from HST data and they
have strong Fe II and Mg II lines in the Sloan spectra. Given that
these systems were chosen because of strong metal lines, they are
probably predisposed to be more metal rich. We have acquired
high-resolution observations of the most promising systems in order to
determine the exact velocity profiles and to remove blends of lines of
different elements (e.g. contribution of Mg I $\lambda$ 2026.5 from
that of Zn II $\lambda$ 2026.1, and of Zn II $\lambda$ 2062.7 from Cr
II $\lambda$ 2062.2). From this sample, a super-solar sub-DLA was
discovered (Khare et al. 2004). Our high-resolution follow-up study of
this sight line has shown that it is the most metal-rich DLA/sub-DLA
known to date (P\'eroux et al. 2006).

The new sample presented here consists of 3 DLAs and 1 sub-DLA at $0.6
< z < 0.9$.  The data were acquired with the Ultraviolet and Visual
Echelle Spectrograph (UVES) mounted on Kueyen Unit 2 Very Large
Telescope (D'Odorico \e\ 2000). The observations were performed in
Service Mode in period 74 between October 2004 and March 2005 under
program number ESO 74.A-0597. We used a combination of 346$+$564 and
390$+$564 nm central wavelength settings appropriate to the range of
wavelengths of the lines we are seeking. The total exposure time for
each object was split into two or three equal observing blocks to
minimize the effect of cosmic rays. A 2$\times$2 CCD binning was used
all along. The properties of the targets and the dates of observations
are summarised in Table~\ref{t:JoO}. The names of the targets are
abbreviated throughout the paper. The full names are respectively:
SDSS J013405.75$+$005109.4, SDSS J025607.24$+$011038.6, SDSS
J110729.03$+$004811.1 and SDSS J232820.37$+$002238.2.

The data were reduced using the most recent version of the UVES
pipeline in {\tt MIDAS} (uves/2.1.0 flmidas/1.1.0). Master bias and
flat images were constructed using calibration frames taken closest in
time to the science frames. The science frames were extracted with the
``optimal'' option. The spectrum was then corrected to vacuum
heliocentric reference. The resulting spectra were combined weighting
each spectrum with its signal-to-noise ratio. The resulting spectral
resolution is R=$\lambda$/$\Delta \lambda \sim$45000. To perform the
abundance studies, the spectra were divided into 50-75 {\AA} regions,
and each region was normalised using cubic spline functions of orders
1 to 5 as the local continuum.

\section{Analysis}

The column densities were estimated by fitting multi-component Voigt
profiles to the observed absorption lines using the program FITS6P
(Welty, Hobbs, \& York 1991) which evolved from the code used by
Vidal-Madjar et al. (1977). FITS6P minimizes the $\chi^{2}$ between
the data and the theoretical Voigt profiles convolved with the
instrumental profile. On some occasions, {\tt fitlyman} in {\tt MIDAS}
(Fontana \& Ballester 1995) was used to allow for simultaneous fits of
several lines of the same ions (i.e. Fe II). The fits were performed
assuming that metal species with similar ionisation potentials can be
fitted using identical velocities and Doppler $b$ parameters. The
redshifts are measured from the Sloan spectra (York et al. 2005). The
atomic data were adopted from Morton (2003).

For the fitting procedure, the same prescription was used for all the
systems. The velocity and Doppler $b$ parameters of the various
components were estimated from the Mg I, Mg II, and Fe II lines and
then kept fixed for the remaining lines, allowing for variations from
one metal species to another in $N$ only. In all cases, the Mg I
$\lambda$ 2026.5 contribution to the Zn II $\lambda$ 2026.1 line was
estimated using the component parameters for Mg I derived from the Mg
I $\lambda$ 2852 profile. For the four quasar absorbers under study,
the Mg II $\lambda \lambda$ 2796 and 2803 profiles were fitted
together, but provide only a lower limit to Mg II owing to saturation
in the central components. In Tables~\ref{t:Q0134} to \ref{t:Q2328},
the resulting column densities in the few weak components that could
not be well-constrained due to noise are marked with '--'; their
contributions to the total column densities are negligible and
therefore not included in the total sum. These are nevertheless listed
for consistency with the velocity structure of the strong lines of Fe
II and Mg II. We also report rest-frame equivalent widths (hereafter
EW) for Mg I, Mg II, Zn II, Cr II and Ca II together with their
1$\sigma$ photon noise uncertainties. York et al. (2006) suggest that
Mg I $\lambda$ 2026 is unimportant to the Zn II $\lambda$ 2026
analysis if EW(Mg I $\lambda$ 2852)$<$0.60\AA. Above this threshold,
Mg I $\lambda$ 2026 competes with Zn II $\lambda$ 2026 or could even
overwhelm it. But below that, it appears that the individual
components that make up strong Mg I lines are not saturated. Since the
ratio of oscillator strengths for the two Mg I transitions is 32, an
unsaturated Mg I component seen in $\lambda$ 2852 would be difficult
to detect at $\lambda$ 2026. We verify this result in several
particular cases at high resolution. We have also checked the Sloan
spectra of our objects for the presence of systems that might contain
lines that would blend with the lines which are being analysed
here. A catalogue of the Sloan systems is being created (York et al.,
2005) where many lines have been identified and double checked with
our high-resolution spectra.

\subsection{SDSS J0134$+$0051, \zem=1.522, \zabs=0.842}

\begin{table*}
\begin{center}
\caption{Parameter fit to the $z_{\rm abs} = 0.842$ SDSS J0134$+$0051 
sub-DLA line. Velocities and $b$ values are in km s$^{-1}$ and column
densities are in cm$^{-2}$. The $\sigma$ values below the column
densities are the error estimates on the column densities.}
\label{t:Q0134}
\begin{tabular}{l r r r r r r r r }
\hline\hline
         &Vel      & b    &\ion{Mg}{i}   &\ion{Mg}{ii}   &\ion{Fe}{ii}   &\ion{Zn}{ii}   &\ion{Cr}{ii}   &\ion{Mn}{ii} \\ 
\hline
N(X)	 &$-$88.5    & 9.9 &--        &--         &--            &4.38E+11   &--              &--        \\
$\sigma$ &    ...    & ... &--        &--         &--         	 &3.72E+11   &--              &--         \\
N(X)     &$-$82.6    & 7.4 &--        & 1.87E+12  &8.17E+11	 &--         &5.38E+12        &--         \\
$\sigma$ &    ...    & ... &--        & 1.13E+11  &2.63E+11	 &--         &1.94E+12        &--         \\
N(X)     &$-$69.9    & 5.3 &5.56E+10  & 1.28E+12  &4.28E+11	 &2.90E+11   &--              &--         \\
$\sigma$ &    ...    & ... &2.32E+10  & 9.70E+10  &1.40E+11	 &1.84E+11   &--              &--         \\
N(X)     &$-$46.5    & 9.1 &--        & 4.48E+12  &3.21E+12	 &5.35E+11   &--              &--         \\
$\sigma$ &    ...    & ... &--        & 1.72E+11  &2.37E+11	 &2.36E+11   &--              &--         \\
N(X)     &$-$31.3    & 6.9 &1.26E+11  & 8.47E+12  &1.25E+13	 &--         &--              & 2.84E+11  \\
$\sigma$ &    ...    & ... &3.10E+10  & 4.68E+11  &8.60E+11	 &--         &--              & 1.24E+11  \\
N(X)     &$-$20.6    & 7.7 &1.37E+11  &$>$1.33E+13&7.10E+13	 &--         &--              &--         \\
$\sigma$ &    ...    & ... &3.68E+10  & ...       &5.32E+12	 &--         &--              &--         \\
N(X)     &$-$10.5    & 6.8 &2.35E+11  &$>$1.44E+13&3.01E+13	 &--         &--              & 4.27E+11  \\
$\sigma$ &    ...    & ... &4.13E+10  & ...       &2.84E+12	 &--         &--              & 1.51E+11  \\
N(X)     &    0.7    &11.5 &--        &$>$1.11E+13&--            &2.14E+11   &--              &--         \\
$\sigma$ &    ...    & ... &--        & ...       &--            &3.83E+11   &--              &--         \\
N(X)     &   14.8    &12.2 &6.14E+11  &$>$1.44E+13&8.93E+13	 &--         &--              & 9.03E+11  \\
$\sigma$ &    ...    & ... &8.02E+10  & ...       &7.35E+12	 &--         &--              & 2.66E+11  \\
N(X)     &   26.3    &10.9 &3.98E+11  &$>$1.36E+13&6.00E+13	 &--         &--              & 9.12E+11  \\
$\sigma$ &    ...    & ... &6.65E+10  & ...       &6.17E+12	 &--         &--              & 2.27E+11  \\
N(X)     &   40.1    & 8.1 &6.00E+10  & 1.61E+13  &2.53E+13	 &--         &--              &--         \\
$\sigma$ &    ...    & ... &3.37E+10  & 8.04E+11  &1.40E+12	 &--         &--              &--         \\
N(X)     &   59.7    & 7.6 &9.14E+10  & 5.66E+12  &6.57E+12	 &--         &--              & 2.02E+11  \\
$\sigma$ &    ...    & ... &2.82E+10  & 2.01E+11  &2.95E+11	 &--         &--              & 1.19E+11  \\
N(X)     &   72.0    & 5.9 &--        &8.89E+11   &--            &--         &1.07E+12        &--         \\
$\sigma$ &    ...    & ... &--        & 1.02E+11  &--         	 &--         &9.79E+11        &--         \\
\hline 				       			 	
\hline 				       			 	
\end{tabular}			       			 	
\end{center}			       			 	
\end{table*}

\begin{figure}
\begin{center}
\includegraphics[height=8cm, width=10.5cm, angle=-90]{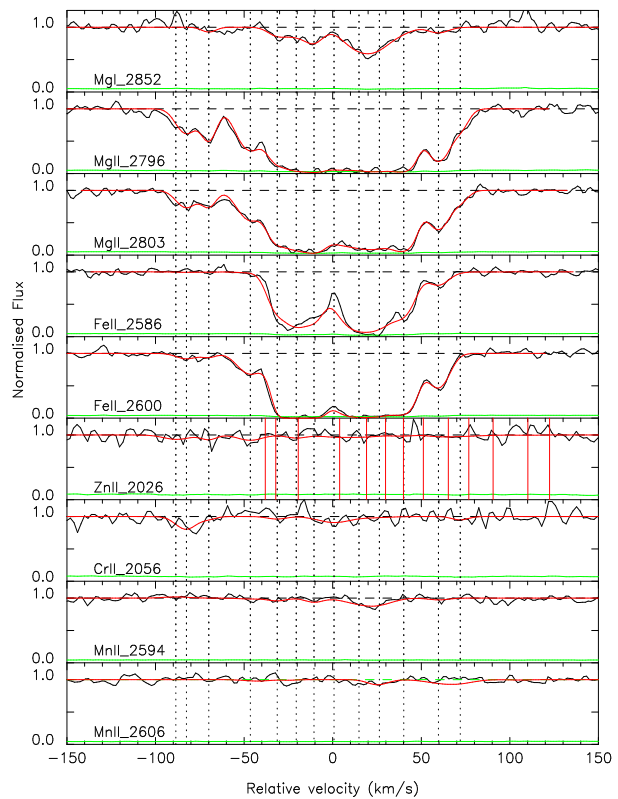}
\caption{Velocity plots of the various lines covered at $z_{\rm
abs}=0.842$ toward SDSS J0134$+$0051. The dotted vertical lines
indicate positions of the components while the solid vertical lines in
the Zn II $\lambda$ 2026 panel indicate the position of the Mg I
$\lambda$ 2026 components. The solid grey line near the zero flux
level is the error array.\label{f:Q0134_others}}
\end{center}
\end{figure}

Voigt profile fitting to the Ly-$\alpha$ absorption line in the HST
STIS spectra for this object (program 9382, PI: Rao) gives log \nhi =
19.93 for this absorber (Sandhya Rao, {\it private communication}). In
line with other \nhi\ measurements reported here, the error estimate
for the column density of this system is derived by moving the
continuum level above and below the best-fit continuum and re-doing
the Voigt profile fitting using the {\tt fitlyman} package within {\tt
MIDAS} (Fontana \& Ballester 1995). This gives log \nhi =
19.93$^{+0.10}_{-0.15}$.

At z$_{\rm abs}$=0.842, Mg I $\lambda$ 2852, Mg II $\lambda \lambda$
2796 and 2803, Fe II $\lambda \lambda$ 2586 and 2600, Zn II $\lambda
\lambda$ 2026 2056, Cr II $\lambda \lambda$ 2056 and 2062 and Mn II
$\lambda \lambda$ 2594 and 2606 are covered. The absorption profiles
show a complex velocity structure with a total of 13 components needed
to properly fit the observed profile. Figure~\ref{f:Q0134_others}
presents the resulting fits and the parameters are given in
Table~\ref{t:Q0134}. Listed are the $N$ and $\sigma N$ values for each
component of each ion.

The rest-frame EW of these lines are measured to be: EW(Mg I $\lambda$
2852)=0.23$\pm$0.01\AA, EW(Mg II $\lambda$ 2796)=1.15$\pm$0.01\AA\ and
EW(Mg II $\lambda$ 2803)=0.96$\pm$0.01\AA. So the Mg I contribution to
the $\lambda$ 2026 line, shown by the solid vertical lines of the Zn II
$\lambda$ 2026 panel in Figure~\ref{f:Q0134_others}, should be
negligible. The Fe II column densities in the components at extreme
negative velocities were constrained by using the Fe II $\lambda
\lambda$ 2586 and 2600 lines simultaneously, since they are weak
in the other Fe II lines. Zn II, Cr II and the two Mn II are extremely
weak in some components and therefore lead to upper limits on the total
column density estimates.

In addition, this rich quasar spectrum has a strong absorption system
at \zabs=0.618. This system is most likely a DLA or sub-DLA given the
strength of the Fe II $\lambda \lambda$ 2249, 2260, 2344, 2374, 2382
and Ti II $\lambda$ 3384 lines in our spectra. Unfortunately, its
Ly-$\alpha$ absorption line lies in a region of the HST spectrum where
the quasar flux is very low owing to the Lyman limit absorption from
another system at $z = 1.272$. Thus it is not possible to estimate the
\nhi\ for the $z=0.618$ system.  Based on Ly-$\alpha$ line for the $z
= 1.272$ system, we estimate that it has log \nhi\ in the range 18.9
to 19.1. This other system at \zabs=1.272 shows very strong C IV and
Mg II doublets, Si II $\lambda$ 1526, Al II $\lambda$ 1670, Mg I $\lambda$
2852, Fe II $\lambda \lambda$ 1608, 2374, 2382, 2600. Also another
absorber is detected at \zabs=1.449 with strong Si IV doublet, Si II
$\lambda$ 1526, C II $\lambda$ 1334, Ni II $\lambda$ 1370, C IV doublet
and Fe II $\lambda$ 2600. 

\subsection{SDSS J0256$+$0110, \zem=1.348, \zabs=0.725}

\begin{table*}
\begin{center}
\caption{Parameter fit to the $z_{\rm abs} = 0.725$ SDSS J0256$+$0110 
DLA line. Velocities and $b$ values are in km s$^{-1}$ and column
densities are in cm$^{-2}$. The $\sigma$ values below the column
densities are the error estimates on the column densities.}
\label{t:Q0256}
\begin{tabular}{l r r r r r r r r r}
\hline\hline
         &Vel      & b    &\ion{Mg}{i}   &\ion{Mg}{ii}   &\ion{Fe}{ii}   &\ion{Zn}{ii}   &\ion{Cr}{ii}   &\ion{Mn}{ii} &\ion{Ti}{ii} \\ 
\hline
N(X)     &$-$140.9& 8.9 &--         &--         &1.02E+10  &--          &5.38E+11   &3.94E+11   &--         \\
$\sigma$ &  ...   &  ...&--         &--         &1.25E+09  &--          &1.57E+12   &1.20E+11   &--         \\
N(X)     &$-$118.7& 8.5 &4.66E+10   &9.33E+11   &5.49E+13  &--          &3.23E+12   &--         &--         \\
$\sigma$ &  ...   &  ...&2.07E+10   &1.68E+11   &1.28E+12  &--          &1.65E+12   &--         &--         \\
N(X)     &$-$98.9 & 11.1&4.77E+10   &2.14E+13   &7.41E+10  &--          &--         &--         &1.51E+11\\
$\sigma$ &  ...   &  ...&2.53E+10   &1.11E+12   &1.73E+09  &--          &--         &--         &7.23E+10\\
N(X)	 &$-$81.1 &  9.2&8.48E+11   &$>$4.28E+13&8.51E+11  &5.72E+11    &--         &1.29E+12   &--        \\
$\sigma$ &  ...   &  ...&3.81E+10   &...        &1.98E+10  &3.47E+11    &--         &1.43E+11   &--        \\
N(X)     &$-$47.3 & 17.1&1.81E+12   &$>$1.08E+14&1.02E+14  &2.16E+12    &4.53E+12   &4.66E+12   &--        \\
$\sigma$ &  ...   &  ...&5.66E+10   &...        &2.38E+12  &5.05E+11    &2.30E+12   &2.16E+11   &--        \\
N(X)     &$-$23.8 &  9.4&4.70E+11   &$>$9.41E+12&1.23E+14  &--          &3.71E+12   &5.31E+11   &--        \\
$\sigma$ &  ...   &  ...&3.76E+10   &...        &5.79E+12  &--          &1.89E+12   &1.42E+11   &--        \\
N(X)     & $-1$.2 & 14.0&1.37E+12   &$>$1.27E+14&2.82E+14  &1.33E+12    &8.81E+12   &1.88E+12   &1.32E+11\\
$\sigma$ &  ...   &  ...&4.79E+10   &...        &6.72E+12  &4.44E+11    &2.31E+12   &1.67E+11   &7.84E+10\\
N(X)     & 31.0   & 10.2&9.29E+11   &$>$6.35E+13&5.13E+13  &7.55E+11    &7.90E+12   &1.59E+12   &4.81E+11\\
$\sigma$ &  ...   &  ...&3.99E+10   &...        &7.60E+12  &3.76E+11    &2.03E+12   &1.47E+11   &7.08E+10\\
N(X)     & 52.8   & 10.6&9.93E+11   &$>$1.01E+14&1.70E+14  &3.19E+12    &9.98E+12   &2.50E+12   &3.67E+11\\
$\sigma$ &  ...   &  ...&4.06E+10   &...        &8.01E+12  &4.94E+11    &2.14E+12   &1.61E+11   &7.06E+10\\
N(X)     & 77.8   &  6.5&7.59E+11   &$>$4.91E+13&3.16E+14  &2.08E+12    &2.50E+09   &1.23E+12   &--        \\
$\sigma$ &  ...   &  ...&3.96E+10   &...        &2.26E+13  &4.32E+11    &1.44E+12   &1.25E+11   &--        \\
N(X)     & 97.2   & 13.6&1.14E+12   &8.20E+13   &1.35E+14  &3.49E+12    &8.02E+12   &1.58E+12   &--        \\
$\sigma$ &  ...   &  ...&4.34E+10   &4.57E+12   &3.14E+12  &5.39E+11    &2.31E+12   &1.62E+11   &--        \\
N(X)     &138.4   &  8.6&2.47E+11   &1.12E+13   &3.55E+08  &5.18E+11    &6.28E+12   &--         &2.16E+11\\
$\sigma$ &  ...   &  ...&2.39E+10   &4.21E+11   &1.67E+07  &3.34E+11    &2.15E+12   &--         &6.30E+10\\
N(X)     &157.8   &  9.5&3.13E+10   &2.23E+12   &--        &--          &4.28E+12   &3.15E+11   &2.03E+11\\
$\sigma$ &  ...   &  ...&2.16E+10   &1.73E+11   &--        &--          &2.09E+12   &1.23E+11   &6.60E+10\\
N(X)     &179.6   & 10.5&--         &8.32E+11   &6.92E+12  &--          &--         &--         &1.11E+11\\
$\sigma$ &  ...   &  ...&--         &1.49E+11   &1.61E+11  &--          &--         &--         &6.78E+10\\
N(X)     &207.0   & 11.6&--         &1.11E+13   &6.16E+12  &--          &--         &2.07E+11   &--        \\
$\sigma$ &  ...   &  ...&--         &3.55E+11   &1.40E+11  &--          &--         &1.32E+11   &--        \\
N(X)     &236.1   & 10.8&1.02E+11   &4.61E+12   &--        &--          &--         &--         &9.07E+10\\
$\sigma$ &  ...   &  ...&2.35E+10   &2.14E+11   &--        &--          &--         &--         &6.82E+10\\
N(X)     &256.3   &  7.6&2.72E+10   &2.22E+12   &5.89E+13  &3.68E+11    &2.97E+12   &1.36E+11   &--        \\
$\sigma$ &  ...   &  ...&1.93E+10   &1.67E+11   &1.37E+12  &3.03E+11    &1.54E+12   &1.09E+11   &--        \\
N(X)     &274.8   &  6.4&2.66E+11   &2.68E+13   &3.55E+13  &3.39E+11    &3.79E+12   &5.88E+11   &--        \\
$\sigma$ &  ...   &  ...&2.25E+10   &2.05E+12   &1.67E+12  &2.83E+11    &1.50E+12   &1.08E+11   &--        \\
N(X)     &289.3   &  5.3&3.62E+11   &1.39E+13   &--        &6.58E+11    &--         &6.57E+11   &9.03E+10\\
$\sigma$ &  ...   &  ...&2.42E+10   &9.57E+11   &--        &2.85E+11    &--         &1.03E+11   &4.91E+10\\
\hline 				       			 	 
\hline 				       			 	 
\end{tabular}			       			 	 
\end{center}			       			 	 
\end{table*}

\begin{figure}
\begin{center}
\includegraphics[height=8cm, width=8cm, angle=-90]{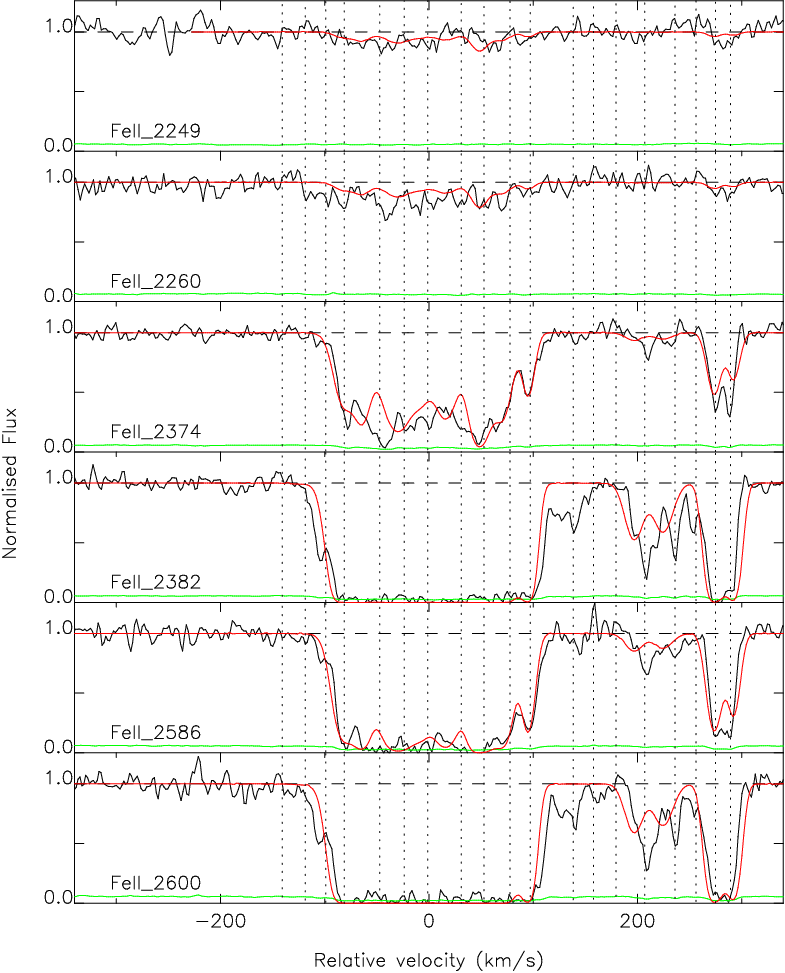}
\vskip 0.2in
\caption{Velocity plots of the Fe II lines detected at $z_{\rm
abs}=0.725$ toward SDSS J0256$+$0110. The dotted lines indicate
positions of the components. The solid grey line near the zero flux
level is the error array.\label{f:Q0256_Fe}} 
\end{center}

\end{figure}
\begin{figure}
\begin{center}
\includegraphics[height=8cm, width=10.5cm, angle=-90]{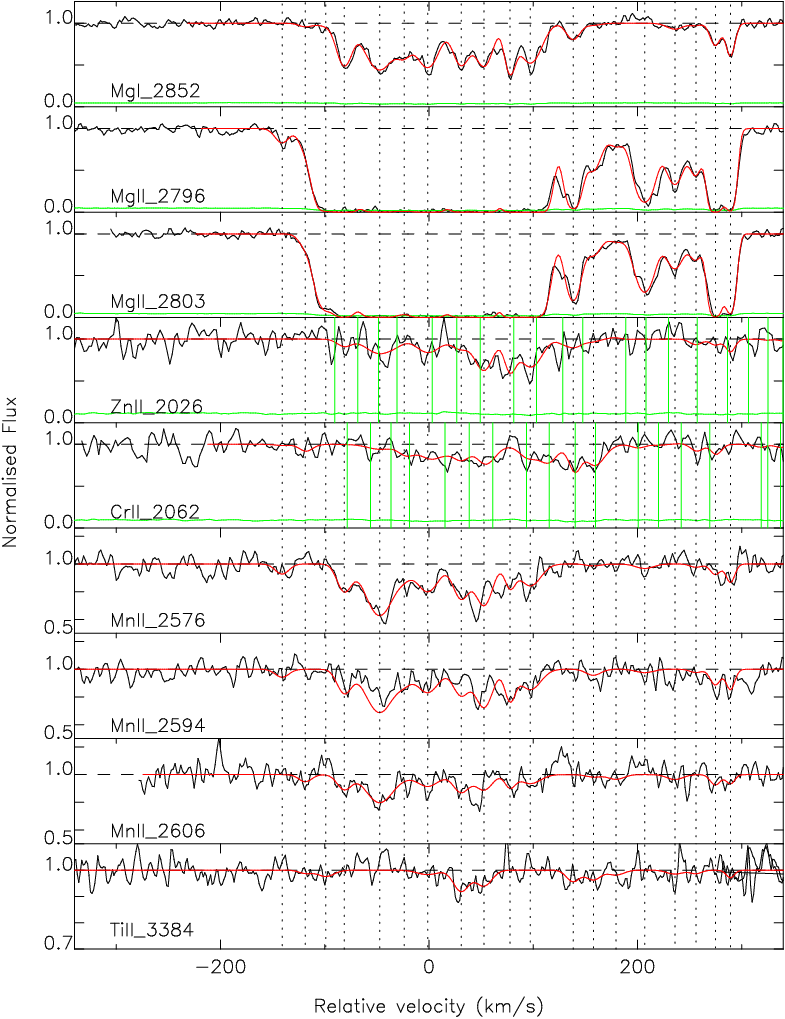}
\vskip 0.2in
\caption{Velocity plots of the lines other than those of Fe II detected at 
$z_{\rm abs}=0.725$ toward SDSS J0256$+$0110. The dotted vertical
lines indicate positions of the components while the solid vertical
lines in the Zn II $\lambda$ 2026 and Cr II $\lambda$ 2062 panels
indicate the position of the Mg I $\lambda$ 2026 and Zn II $\lambda$
2062 components respectively. The solid grey line near the zero flux
level is the error array.\label{f:Q0256_others}}
\end{center}
\end{figure}

Voigt profile fitting to the Ly-$\alpha$ absorption line in the HST
STIS spectra for this object (Rao, Turnshek \& Nestor, 2006) give log
\nhi = 20.70$^{+0.11}_{-0.22}$. Mg I $\lambda$ 2852, Mg II $\lambda
\lambda$ 2796 and 2803, Fe II $\lambda \lambda$ 2249, 2260, 2374, 2382,
2586 and 2600, Zn II $\lambda$ 2026, Zn II$+$Cr II $\lambda$ 2062, Mn II
$\lambda \lambda$ 2576, 2594 and 2606 and Ti II $\lambda$ 3384 are
detected in this system. The absorption profiles show a complex
velocity structure with a total of 19 components needed to best fit
the profile. The results of the profile fitting analysis are
summarised in Table~\ref{t:Q0256}, while Figure~\ref{f:Q0256_Fe} and
Figure~\ref{f:Q0256_others} illustrate the fits. Strong C IV is also
detected at \zabs=1.326 with associated Fe II $\lambda$ 1608 in the
same UVES spectrum.

At \zabs=0.725, the rest-frame EW of these lines are measured to be:
EW(Mg I $\lambda$ 2852)=0.94$\pm$0.01\AA, EW(Mg II $\lambda$
2796)=3.22$\pm$0.01\AA, EW(Mg II $\lambda$ 2803)=2.88$\pm$0.01\AA,
EW(Zn II $\lambda$ 2026)=0.24$\pm$0.02\AA\ and EW(Zn II$+$Cr II $\lambda$
2062)=0.23$\pm$0.02\AA\ while Cr II $\lambda$ 2056 is not detected. We
have carefully checked that these detections/non-detections are
self-consistent considering the respective oscillator strengths and
noise in the appropriate region. In this system, the Mg I $\lambda$
2852 line provides a good guide to the component decomposition. Many
Fe II lines are detected; the weakest ones are used to constrain the
most central components of the profile while the strongest lines are
used to accurately determine the column density in the weaker
components. The component at $-$1.2 \kms\ seems broader in Mg I
$\lambda$ 2852 than in Fe II $\lambda$ 2374. Similarly, the saturated
Fe II lines show a very weak extra component at v=$-$140.9 \kms\ not
seen in the other elements. Zn II$+$Mg I $\lambda$ 2026 and Zn II$+$Cr II
$\lambda$ 2062 are clearly detected and lead to robust column density
determinations. The solid vertical lines in the Zn II $\lambda$ 2026
and Cr II $\lambda$ 2062 panels of Figure~\ref{f:Q0256_others} indicate
the position of the Mg I $\lambda$ 2026 and Zn II $\lambda$ 2062
components respectively. Mn II and Ti II are also unambiguously detected
and accurate column densities could be determined.

\subsection{SDSS J1107$+$0048, \zem=1.388, \zabs=0.740}

\begin{table*}
\begin{center}
\caption{Parameter fit to the $z_{\rm abs} = 0.740$ SDSS J1107$+$0048 
DLA line. Velocities and $b$ values are in km s$^{-1}$ and column
densities are in cm$^{-2}$. The $\sigma$ values below the column
densities are the error estimates on the column densities.}
\label{t:Q1107}
\begin{tabular}{l r r r r r r r r}
\hline\hline
         &Vel      & b    &\ion{Mg}{i}   &\ion{Mg}{ii}   &\ion{Fe}{ii}   &\ion{Zn}{ii}   &\ion{Cr}{ii}   &\ion{Mn}{ii} \\ 
\hline
N(X)     &$-$186.8 & 6.8 &6.22E+10   &9.15E+12    &1.17E+13   &--          &--           &--      \\
$\sigma$ &     ... & ... &1.64E+10   &3.97E+11    &3.39E+12   &--          &--	         &--      \\
N(X)     &$-1$66.6 & 6.6 &1.21E+11   &1.19E+13    &--         &--          &8.0992E+12   &3.60E+11\\
$\sigma$ &     ... & ... &1.74E+10   &6.34E+11    &--         &--          &3.1904E+12   &2.11E+11\\
N(X)	 &$-$150.6 & 7.2 &1.97E+11   &1.25E+13    &1.70E+13   &--          &--           &--      \\
$\sigma$ &     ... & ... &1.97E+10   &6.08E+11    &2.97E+12   &--          &--           &--      \\
N(X)     &$-$112.5 & 6.4 &4.99E+10   &1.90E+13    &8.91E+12   &--          &--           &6.59E+11\\
$\sigma$ &     ... & ... &1.60E+10   &1.36E+12    &2.57E+12   &--          &--           &2.20E+11\\
N(X)     & $-$88.6 &12.0 &7.24E+11   &$>$6.31E+13 &5.62E+14   &2.45E+12    &1.1855E+13   &2.33E+12\\
$\sigma$ &     ... & ... &3.03E+10   &...         &2.65E+13   &7.55E+11    &4.0125E+12   &3.30E+11\\
N(X)     & $-$66.0 &10.8 &4.05E+11   &$>$5.03E+13 &2.04E+14   &1.35E+12    &4.4839E+12   &6.49E+11\\
$\sigma$ &     ... & ... &2.54E+10   &...         &9.63E+12   &6.60E+11    &3.3508E+12   &2.74E+11\\
N(X)     & $-$38.9 & 9.7 &3.33E+11   &$>$5.27E+13 &1.51E+14   &9.92E+11    &--           &--      \\
$\sigma$ &     ... & ... &2.31E+10   &...         &7.13E+12   &6.09E+11    &--           &--      \\
N(X)     & $-$17.9 & 9.3 &3.66E+11   &$>$4.59E+13 &9.12E+13   &1.41E+12    &--           &4.77E+11\\
$\sigma$ &     ... & ... &2.43E+10   &...         &4.30E+12   &6.37E+11    &--           &2.53E+11\\
N(X)     &     6.4 &11.7 &1.66E+12   &$>$4.84E+13 &1.02E+15   &2.18E+12    &1.4433E+13   &5.61E+12\\
$\sigma$ &     ... & ... &5.18E+10   &...         &4.82E+13   &7.44E+11    &4.3565E+12   &4.57E+11\\
N(X)     &    25.3 & 9.0 &8.11E+11   &$>$3.97E+13 &1.00E+14   &--          &8.4768E+12   &1.05E+12\\
$\sigma$ &     ... & ... &4.13E+10   &...         &9.65E+12   &--          &3.8171E+12   &3.06E+11\\
N(X)     &    45.0 &10.3 &2.26E+12   &$>$5.34E+13 &7.94E+14   &2.35E+12    &1.4673E+13   &5.03E+12\\
$\sigma$ &     ... & ... &7.72E+10   &...         &3.74E+13   &7.37E+11    &4.2857E+12   &4.29E+11\\
N(X)     &    66.2 & 9.9 &1.01E+12   &$>$2.26E+13 &1.86E+14   &--          &7.1532E+12   &4.73E+11\\
$\sigma$ &     ... & ... &4.49E+10   &...         &1.33E+13   &--          &3.7389E+12   &2.97E+11\\
N(X)     &    82.6 &12.4 &7.07E+11   &$>$7.75E+13 &1.86E+14   &--          &--           &2.07E+12\\
$\sigma$ &     ... & ... &3.46E+10   &...         &8.77e+12   &--          &--           &3.46E+11\\
N(X)     &   110.3 & 7.7 &5.63E+10   &3.08E+13    &6.16E+13   &--          &--           &--          \\
$\sigma$ &     ... & ... &1.73E+10   &2.39E+12    &2.91E+12   &--          &--           &--          \\
N(X)     &   129.6 & 8.3 &--         &2.05E +12   &--         &7.32E+11    &3.1583E+12   &--          \\
$\sigma$ &     ... & ... &--         &1.75E+11    &--         &5.52E+11    &2.3436E+12   &--           \\
\hline 				       			 	 
\hline 				       			 	 
\end{tabular}			       			 	 
\end{center}			       			 	 
\end{table*}

\begin{figure}
\begin{center}
\includegraphics[height=8cm, width=8cm, angle=-90]{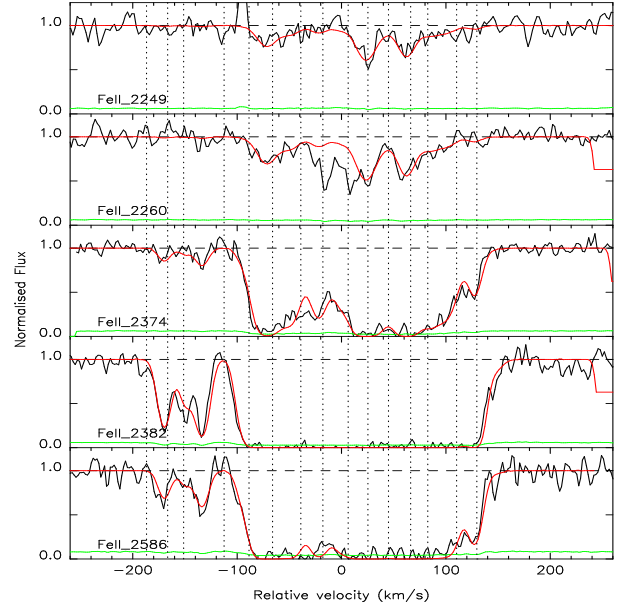}
\caption{Velocity plots of the Fe II lines detected at $z_{\rm
abs}=0.740$ toward SDSS J1107$+$0048. The dotted lines indicate
positions of the components. The components at $-$17.9 and $+$6.4 km/s
in Fe II $\lambda$ 2260 are clearly blended with
interlopers. The solid grey line near the zero flux
level is the error array.\label{f:Q1107_Fe}}
\end{center}
\end{figure}

\begin{figure}
\begin{center}
\includegraphics[height=8cm, width=8cm, angle=-90]{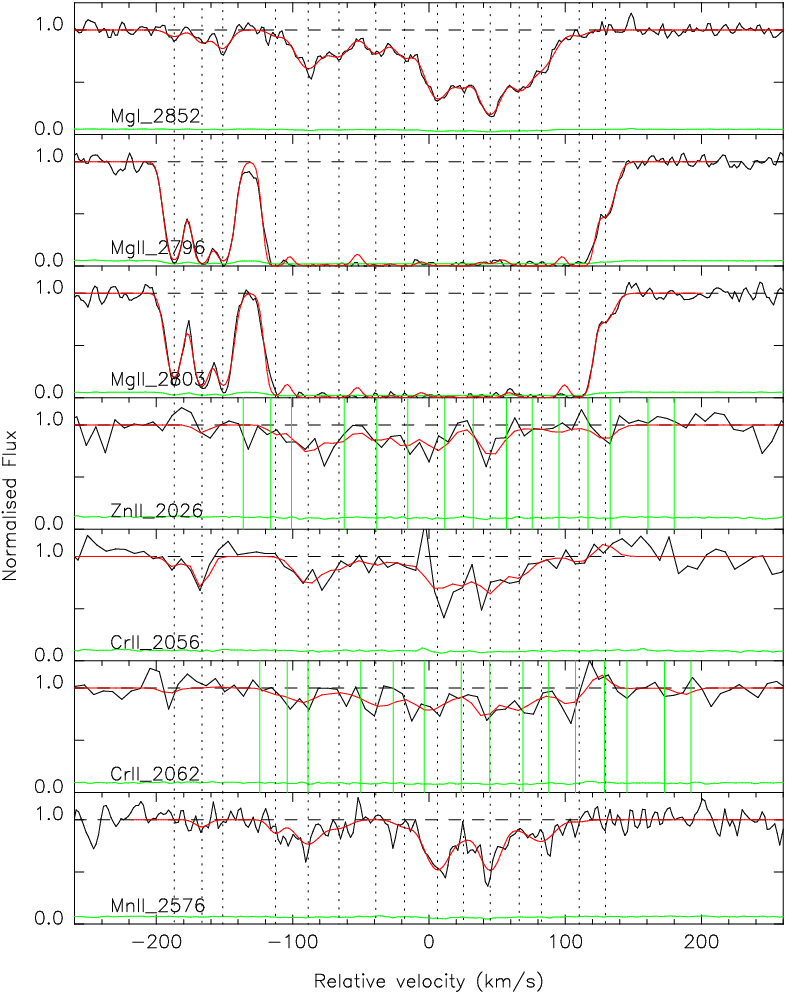}
\caption{Velocity plots of the lines other than those of Fe II detected 
at $z_{\rm abs}=0.740$ toward SDSS J1107$+$0048. The dotted vertical
lines indicate positions of the components while the solid vertical
lines in the Zn II $\lambda$ 2026 and Cr II $\lambda$ 2062 panels
indicate the position of the Mg I $\lambda$ 2026 and Zn II $\lambda$
2062 components respectively. The solid grey line near the zero flux
level is the error array.\label{f:Q1107_others}}
\end{center}
\end{figure}

Voigt profile fitting to the Ly-$\alpha$ absorption line in the HST
STIS spectra for this object (Rao, Turnshek \& Nestor, 2006) give log
\nhi = 21.00$^{+0.02}_{-0,05}$. This system was first studied by Khare
et al. (2004) who reported abundance determination for various
elements. In the new high-resolution data, several lines are detected
in this system including Mg I $\lambda$ 2852, Mg II $\lambda \lambda$
2796 and 2803, Fe II $\lambda \lambda$ 2249, 2260, 2374, 2382 and
2586, Zn II $\lambda$ 2026, Cr II $\lambda \lambda$ 2056 and 2062 and
Mn II $\lambda$ 2576. The velocity profile is best fitted with 15
components. Figure~\ref{f:Q1107_Fe} and Figure~\ref{f:Q1107_others}
present the resulting fits and the parameters deduced from these fits
are listed in Table~\ref{t:Q1107}. A strong C IV doublet is also
detected at \zabs=1.369 as well as a strong Mg II doublet at
\zabs=1.070 in the same UVES spectrum.

Based on the profile fitting results for Mg I $\lambda$ 2852, the EW
is estimated to be EW(Mg I $\lambda$ 2852)=0.83$\pm$0.02\AA, while
EW(Mg II $\lambda$ 2796)=2.74$\pm$0.01\AA, EW(Mg II $\lambda$
2803)=2.62$\pm$0.01\AA, EW(Zn II $\lambda$ 2026)=0.19$\pm$0.03\AA,
EW(Cr II $\lambda$ 2056)=0.13$\pm$0.05\AA\ and EW(Cr II$+$Zn II
$\lambda$ 2062)=0.13$\pm$0.05\AA. The broad $\lambda$ 2852 absorption
from 0--100 \kms\ in the top panel of Figure~\ref{f:Q1107_others}
would appear in the $\lambda$ 2026 line of Mg I between 50 and 150
\kms\ of the Mg I $\lambda$ 2026 panel (4$^{th}$ from top) as
indicated by the solid vertical lines. However, the profile is
remarkably free of attenuation in that velocity range. The feature at
114 \kms\ in Zn II $\lambda$ 2026 may contain a shadow of the deepest
part of the $\lambda$ 2852 profile. Nevertheless, we can conclude that
the contribution of Mg I to the Zn II $\lambda$ 2026 line is
negligible.

Mg I $\lambda$ 2852 is clearly detected in this system and provides a
good guidance to the velocity profile of the system. The 5 detected Fe
II lines vary considerably in terms of oscillator strengths and
therefore provide a fairly good determination of the total column
density albeit with large error bars. In particular, the components at
$-$17.9 and $+$6.4 \kms\ in Fe II $\lambda$ 2260 are clearly blended
with interlopers. Zn II and Cr II are both detected providing clear
column density measurements. The determination of the column density
of Zn II is based on Zn II $\lambda$ 2026 while Cr II is based on the
$\lambda$ 2056 transition.

Finally, lines of Mn II are also present and give an accurate
determination of the oscillator strength, $f$, and Mn column
density. The resulting column densities are in excellent agreement
with values derived by Khare et al. (2004) from MMT data (see
Table~\ref{t:comp} and discussion in Section 4.1). While the MMT data
have higher signal-to-noise ratio, only the high resolution of the
UVES observations allows one to disentangle the possible blending of
lines as for example for Zn II $\lambda$ 2026.1 and Mg I $\lambda$
2026.5 as in this system.

\subsection{SDSS J2328$+$0022, \zem=1.309, \zabs=0.652}     

\begin{table*}
\begin{center}
\caption{Parameter fit to the $z_{\rm abs} = 0.652$ SDSS J2328$+$0022 
DLA line. Velocities and $b$ values are in km s$^{-1}$ and column
densities are in cm$^{-2}$. The $\sigma$ values below the column
densities are the error estimates on the column densities.}
\label{t:Q2328}
\begin{tabular}{l r r r r r r r r r}
\hline\hline
         &Vel      & b    &\ion{Mg}{i}   &\ion{Mg}{ii}   &\ion{Fe}{ii}   &\ion{Zn}{ii}   &\ion{Cr}{ii}   &\ion{Mn}{ii} &\ion{Ca}{ii}\\ 
\hline
N(X)     &$-$82.3& 3.1&4.54E+10   &$>$1.09E+12&4.57E+11   &--         &1.8532E+12   &--           &--         \\
$\sigma$ &    ...& ...&1.06E+10   &...        &3.75E+11   &--         &1.1182E+12   &--           &--         \\
N(X)     &$-$59.8& 3.7&--         &   4.01E+11&--         &--         &1.8929E+12   &1.43E+11     &3.45E+10   \\
$\sigma$ &    ...& ...&--         &   1.02E+11&--         &--         &9.2035E+11   &6.49E+10     &1.76E+10   \\
N(X)     &$-$40.7&12.3&--         &   1.16E+13&5.75E+12   &--         &3.3529E+12   &7.78E+11     &--         \\
$\sigma$ &    ...& ...&--         &   4.90E+11&1.34E+11   &--         &1.8466E+12   &1.00E+11     &--         \\
N(X)	 &$-$20.6& 8.8&1.27E+12   &$>$2.97E+14&5.37E+13   &--         &1.5493E+12   &--           &2.98E+11   \\
$\sigma$ &    ...& ...&3.68E+10   &...        &1.25E+12   &--         &1.5481E+12   &--           &2.89E+10   \\
N(X)     & $-$1.6& 6.0&1.08E+12   &$>$1.72E+13&1.44E+14   &1.52E+12   &--           &1.81E+12     &6.02E+11   \\
$\sigma$ &    ...& ...&4.28E+10   &...        &3.38E+12   &9.86E+11   &--           &1.04E+11     &2.94E+10   \\
N(X)     &   15.3& 8.7&2.44E+12   &$>$1.23E+14&2.51E+14   &--         &4.0513E+12   &1.90E+12     &1.17E+12   \\
$\sigma$ &    ...& ...&8.14E+10   &...        &5.85E+12   &--         &1.6278E+12   &1.13E+11     &3.84E+10   \\
N(X)     &   35.3& 9.0&1.06E+12   &$>$2.05E+13&1.20E+14   &--         &1.5626E+12   &9.27E+11     &1.62E+11   \\
$\sigma$ &    ...& ...&3.58E+10   &...        &2.80E+12   &--         &1.5491E+12   &1.04E+11     &2.84E+10   \\
N(X)     &   57.6&13.1&7.99E+11   &$>$3.15E+14&1.05E+14   &--         &--           &1.29E+12     &4.28E+11   \\
$\sigma$ &    ...& ...&2.80E+10   &...        &2.44E+12   &--         &--           &1.22E+11     &3.48E+10   \\
N(X)     &   85.5&10.6&4.28E+10   &$>$4.56E+12&3.39E+12   &--         &--           &--           &--         \\
$\sigma$ &    ...& ...&1.72E+10   &...        &1.59E+11   &--         &--           &--           &--         \\
N(X)     &  123.2& 6.1&--         &   5.67E+12&3.16E+12   &--         &1.6275E+12   &9.39E+10     &--         \\
$\sigma$ &    ...& ...&--         &   2.36E+11&1.49E+11   &--         &1.3047E+12   &7.66E+10     &--         \\
N(X)     &  140.6& 4.3&--         &   3.23E+12&1.74E+12   &--         &2.3261E+12   &--           &--         \\
$\sigma$ &    ...& ...&--         &   1.77E+11&1.24E+11   &--         &9.3863E+11   &--           &--         \\
N(X)     &  154.2& 6.2&--         &   5.00E+12&4.07E+12   &1.19E+12   &--           &--           &--         \\
$\sigma$ &    ...& ...&--         &   2.16E+11&4.95E+10   &9.21E+11   &--           &--           &--         \\
N(X)     &  198.1& 3.9&--         &   7.01E+11&3.39E+11   &--         &4.1262E+12   &--           &--         \\
$\sigma$ &    ...& ...&--         &   9.56E+10&9.77E+10   &--         &1.3373E+12   &--           &--         \\
N(X)     &  224.1& 8.1&--         &   1.03E+12&5.37E+11   &--         &--           &--           &--         \\
$\sigma$ &    ...& ...&--         &   1.25E+11&1.24E+11   &--         &--           &--           &--         \\
\hline 				       			 	 
\hline 				       			 	 
\end{tabular}			       			 	 
\end{center}			       			 	 
\end{table*}

\begin{figure}
\begin{center}
\includegraphics[height=8cm, width=8cm, angle=-90]{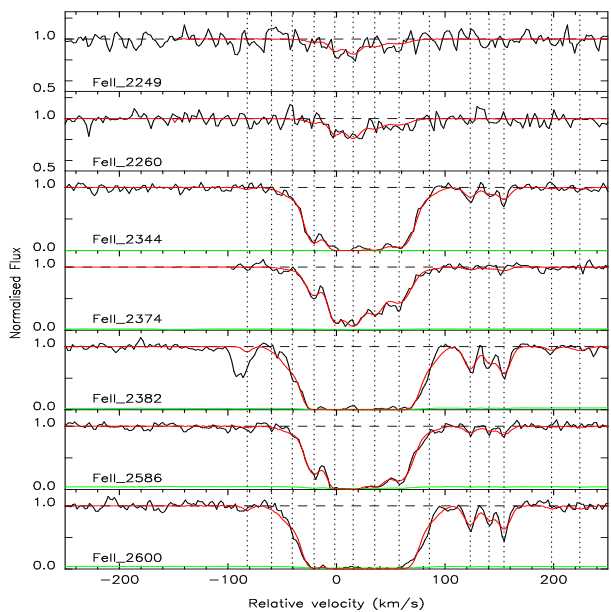}
\caption{Velocity plots of the Fe II lines detected at $z_{\rm
abs}=0.652$ toward SDSS J2328$+$0022. The dotted lines indicate
positions of the components. The solid grey line near the zero flux
level is the error array.\label{f:Q2328_Fe}} 
\end{center}
\end{figure}

\begin{figure}
\begin{center}
\includegraphics[height=8cm, width=10.5cm, angle=-90]{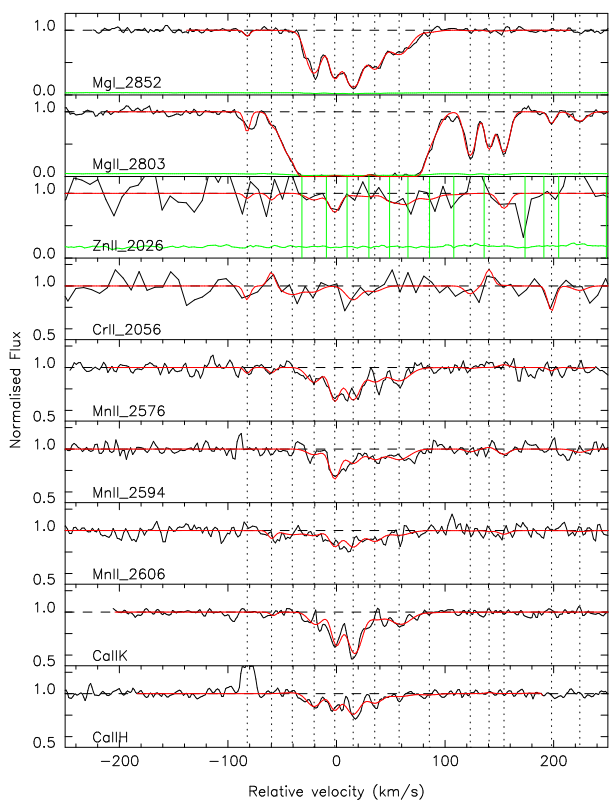}
\caption{Velocity plots of the lines other than those of Fe II detected at 
$z_{\rm abs}=0.652$ toward SDSS J2328$+$0022. The dotted vertical
lines indicate positions of the components while the solid vertical
lines in the Zn II $\lambda$ 2026 panel indicate the position of the
Mg I $\lambda$ 2026 components. The solid grey line near the zero flux
level is the error array.\label{f:Q2328_others}}
\end{center}
\end{figure}

Voigt profile fitting to the Ly-$\alpha$ absorption line in the HST
STIS spectra for this object (Rao, Turnshek \& Nestor, 2006) gives log
\nhi = 20.32$^{+0.06}_{-0.07}$. Several lines are detected in this
system in the UVES spectrum, including Mg I $\lambda$ 2852, Mg II
$\lambda$ 2803 (Mg II $\lambda$ 2796 falls in a spectral gap), Fe II
$\lambda \lambda$ 2249, 2260, 2344, 2374, 2382, 2586 and 2600, Zn II
$\lambda$ 2062, Cr II $\lambda \lambda$ 2056 and 2062, Mn II $\lambda
\lambda$ 2576, 2594 and 2606 and even Ca II K and H $\lambda \lambda$
3933 3969. The absorption profiles show a complex velocity
structure well fitted with a total of 14
components. Figure~\ref{f:Q2328_Fe} and Figure~\ref{f:Q2328_others}
present the resulting fits and Table~\ref{t:Q2328} the associated
parameters. No lines at other absorption redshifts are seen in either
the Sloan or the UVES spectrum of that quasar.

The EW of Zn II and Cr II is bracketed assuming the whole range of
velocity components seen in Mg II (this provides a $<$3$\sigma$ limit)
on one hand and taking just the "innermost region" (which gives a
$>$3$\sigma$ limit) on the other hand. We derive: 0.11$<$EW(Zn II
$\lambda$ 2026)$<$0.12\AA, 0.018$<$EW(Cr II $\lambda$
2056)$<$0.022\AA\ and 0.05$<$EW(Cr II$+$Zn II $\lambda$
2062)$<$0.07\AA. No shadow of Mg I $\lambda$ 2852 is seen in the Zn II
$\lambda$ 2026 frame of Figure~\ref{f:Q2328_others} at the expected
offset position of 50 \kms\ for Mg I $\lambda$ 2026 (solid vertical
lines), once again showing that the components of $\lambda$ 2852 are
not saturated enough to affect our results for Zn II. Indeed, the Mg I
line is clearly detected in this system and leads to a robust
determination of its total column density. The weaker Fe II components
at high positive and negative velocities were constrained using the
$\lambda \lambda$ 2382, 2586 and 2600 lines, since these components
are poorly constrained by the weaker $\lambda \lambda$ 2249 and 2260
lines. Zn II and Cr II are detected at $\lambda
\lambda \lambda$ 2026, 2056 and 2062, respectively.
Mn II is clearly detected in three different lines and leads to a
robust column density determination. The Ca II $\lambda \lambda$ 3933
and 3969 lines are also detected and provide a robust determination of
the column densities. The rest-frame EW of these lines are measured to
be: EW(Mg I $\lambda$ 2852)=0.57$\pm$0.01\AA, EW(Mg II $\lambda$
2803)=1.60$\pm$0.01\AA, EW(Ca II $\lambda$ 3933)=0.20$\pm$0.02\AA\ and
EW(Ca II $\lambda$ 3968)=0.13$\pm$0.01\AA. This a case where the
contribution of Mg I to the Zn II $\lambda$ 2026 line is probably
unimportant.

\section{Results}

\subsection{Total Column Densities}

The resulting total column densities for each element detected in the
four quasar absorbers studied here are summarised in
Table~\ref{t:total_N} together with the associated error bars. In
calculating these total column densities, we used the component
structure from the Fe II and Mg II lines (where all components are
definitely real), but did not include the weak components listed as
'--' in Tables~\ref{t:Q0134} to \ref{t:Q2328}. Mg I, Mg II and Fe II
are clearly detected in all systems, while Zn II and Cr II are only
detected in three of the four systems. For comparison, we also provide
N(Zn II) derived from a Apparent Optical Depth (AOD) method. While
this provides a good consistency check that the contribution of weak
components is really small compared to the contribution from the
'core' components, the AOD leads to an overestimate of the column
densities in the presence of blends (i.e. SDSS J2328$+$0022). Ti II
and Ca II are each detected in one quasar absorber in our sample.

As often observed in high-resolution observations of quasar absorbers,
one can note that the unsaturated lines of Mg II and Fe II appear near
the extremes of the profiles. While this is also the case in our own
Galaxy, it is still difficult to explain if no star formation took
place in these regions.

\begin{table*}
\begin{center}
\caption{Total column densities for each absorber. The numbers in brackets 
correspond to N(Zn II) determination with Apparent Optical Depth (AOD)
method. While this provides a good consistency check that the
contribution of weak components is really small compared to the
contribution from the 'core' components, such method leads to another
estimate of the column densities in the presence of blends (i.e. SDSS
J2328$+$0022). Ti II is detected toward SDSS J0256$+$0110 and Ca II
toward SDSS J2328$+$0022.}
\label{t:total_N}
\begin{tabular}{l r r r r r r r r r}
\hline\hline
& $z_{\rm abs}$ &$\log$ $N_{\rm HI}$ &$\log$ $N_{\rm MgI}$ &$\log$
$N_{\rm MgII}$&$\log$ $N_{\rm FeII}$ &$\log$ $N_{\rm ZnII}$ &$\log$
$N_{\rm CrII}$ &$\log$ $N_{\rm MnII}$ &$\log$ $N_{\rm TiII \hspace{0.05cm} or \hspace{0.05cm} CaII}$ \\
\hline
J0134$+$0051 &0.842  &$19.93^{+0.10}_{-0.15}$ &12.23$\pm$0.04 &$>$14.02  &14.47$\pm$0.01  &$<$12.17\tiny{(11.09)}        &$<$12.81          &$<$12.44        &...\\
J0256$+$0110 &0.725  &$20.70^{+0.11}_{-0.22}$ &12.97$\pm$0.01 &$>$14.83  &15.13$\pm$0.30  &13.19$\pm$0.04\tiny{(13.19)}  &13.81$\pm$0.05    &13.24$\pm$0.02  &12.27$\pm$0.05\\
J1107$+$0048 &0.740  &$21.00^{+0.02}_{-0.05}$ &12.94$\pm$0.01 &$>$14.73  &15.53$\pm$0.02  &13.06$\pm$0.15\tiny{(13.01)}  &13.84$\pm$0.06    &13.27$\pm$0.03  &...\\
J2328$+$0022 &0.652  &$20.32^{+0.06}_{-0.07}$ &12.83$\pm$0.01 &$>$14.91  &14.84$\pm$0.01  &12.43$\pm$0.15\tiny{(13.00)}  &13.35$\pm$0.19    &12.84$\pm$0.02  &12.43$\pm$0.03\\
\hline 				       			 	 
\hline 				       			 	 
\end{tabular}			       			 	 
\end{center}			       			 	 
\end{table*}

\begin{table}
\begin{center}
\caption{Medium/high-resolution column densities comparison.}
\label{t:comp}
\begin{tabular}{l c c c c}
\hline\hline
&\multicolumn{2}{c}{SDSS J1107$+$0048}&\multicolumn{2}{c}{SDSS J0933$+$733}\\
&med res &high res &med res & high res\\
\hline
$\log$ \ion{Mg}{i}   &12.73$^{0.27}_{0.82}$   &12.94$\pm$0.01    &$<$12.75               &...  \\
$\log$ \ion{Mg}{ii}  &...                     &$>$14.73          &...	                 &....  \\
$\log$ \ion{Fe}{ii}  &15.52$^{0.06}_{0.06}$   &15.53$\pm$0.02    &...	                 &15.19$\pm$0.01  \\
$\log$ \ion{Zn}{ii}  &13.03$^{0.05}_{0.05}$   &13.06$\pm$0.15    &12.67$^{0.12}_{0.16}$	 &12.71$\pm$0.02  \\
$\log$ \ion{Cr}{ii}  &13.75$^{0.03}_{0.03}$   &13.84$\pm$0.06    &13.46$^{0.08}_{0.09}$  &13.56$\pm$0.01  \\
$\log$ \ion{Mn}{ii}  &13.34$^{0.04}_{0.07}$   &13.27$\pm$0.03    &...	                 &12.96$\pm$0.01  \\
$\log$ \ion{Ti}{ii}  &$<$12.82                &...	         &12.85$^{0.21}_{0.09}$	 &$<$12.43  \\
$\log$ \ion{Co}{ii}  &$<$13.18	              &...               &$<$13.30	         &$<$12.97  \\
$\log$ \ion{Si}{ii}  &...	              &...               &15.52$^{0.03}_{0.03}$	 &$>$15.55 \\
$\log$ \ion{Ni}{ii}  &...	              &...               &13.91$^{0.01}_{0.01}$	 &13.92$\pm$0.01  \\
\hline 				       			 	 
\hline 				       			 	 
\end{tabular}			       			 	 
\end{center}			       			 	 
\end{table}			       			 	 

Our new high-resolution results allow us to remove any doubts about
anomalous abundances caused by saturation effects. In
Table~\ref{t:comp}, we compare results from medium and high-resolution
in two cases: the DLA with \zabs=0.740, log \nhi=21.00 toward SDSS
J1107$+$0048 presented in this paper and the DLA with \zabs=1.479 log
\nhi=21.62 toward SDSS J0933$+$733 (Khare et al.  2004; Rao,
et al. 2005).  The Multiple Mirror Telescope (MMT) medium-resolution
spectra do not have the high resolution that characterises the UVES
(SDSS J1107$+$0048) and Keck data (SDSS J0933$+$733). Still, in both
cases, we find the column densities to agree well within the errors
(with the exception of Cr II in SDSS J1107$+$0048 and Ti II in SDSS
J0933$+$733). Nevertheless, some Zn values in the literature may be in
error because of having high Mg I and not having verification of the
EW(Mg I $\lambda$ 2852) as an indicator of possible contamination. For
instance e.g. Pettini et al. (1994), Prochaska \& Wolfe (2002), Nestor et
al. (2003), Khare et al. (2004) and Meiring et al. (2006) have
considered Mg I contamination, whereas e.g. Prochaska \& Wolfe (1999) and
Prochaska et al. (2003) have not.

\subsection{Abundances}

The total relative abundances were calculated with respect to solar using
the following formula:

\begin{equation}
[X/H] =\log [N(X)/N(H)]_{DLA}- \log [N(X)/N(H)]_{\odot}
\end{equation}

where $\log [N(X)/N(H)]_{\odot}$ is the solar abundance and is taken
from Asplund et al. (2005) adopting the mean of photospheric and
meteoritic values for Mg, Fe, Zn, Cr, Ti, Ca and the meteoritic value
for Mn. These values are recalled on the top line of
Table~\ref{t:ab}. The abundance in Ca is a lower limit since the
ionization potential of Ca II (11.868 eV) is lower than the one of HI
and therefore Ca II represents only a portion of the total calcium contained
in a neutral gas cloud.

Prochaska (2003) and more recently Rodriguez et al. (2006) studied the
homogeneity across the profile of the quasar absorbers using pixel
analysis for high-redshift absorbers. Both these studies find the
chemical abundances to be rather uniform. Here, we just compare
elemental ratios in components determined by the Voigt profile
fitting. Using the better-determined components in SDSS J0256$+$0110
and SDSS J1107$+$0048, we find that [Fe/Zn] varies by a factor of 5
and 6 respectively. [Fe/Mg] in the weak Mg II components and [Mn/Fe]
in the strong components are found to vary by no more than a factor of
2 or 3 in all systems under study. Our findings are different from
what higher redshifts have shown and might be the signature of
multiple episodes of star formation.

\begin{table*}
\begin{center}
\caption{Abundances with respect to solar, [X/H], using the standard 
definition: $[X/H] = \log [N(X)/N(H)]_{DLA}- \log
[N(X)/N(H)]_{\odot}$. The error bars on [X/H] include both the errors in
log $N(X)$ and \loghi.}
\label{t:ab}
\begin{tabular}{l r r r r r r r r}
\hline\hline
& $z_{\rm abs}$ &\ion{Mg}{} &\ion{Fe}{} &\ion{Zn}{} &\ion{Cr}{}
&\ion{Mn}{} &\ion{Ti}{} &\ion{Ca}{} \\
\hline
A(X/N)$_{\sun}$&--         &$-$4.47 &$-$4.55            &$-$7.40          &$-$6.37          &$-$6.53          &$-$7.11 &$-$5.69\\
\hline
SDSS J0134$+$0051 &0.842   &$>-$1.44 &$-$0.91$\pm$0.16   &$<-$0.36         &$<-$0.75         &$<-$0.96         &...&...\\
SDSS J0256$+$0110 &0.725   &$>-$1.40 &$-$1.02$\pm$0.41   &$-$0.11$\pm$0.04 &$-$0.52$\pm$0.27 &$-$0.93$\pm$0.24 &$-$1.32$\pm$0.28&...\\
SDSS J1107$+$0048 &0.740   &$>-$1.80 &$-$0.92$\pm$0.04   &$-$0.54$\pm$0.20 &$-$0.79$\pm$0.11 &$-$1.20$\pm$0.08 &... &...\\
SDSS J2328$+$0022 &0.652   &$>-$0.94 &$-$0.93$\pm$0.07   &$-$0.49$\pm$0.22 &$-$0.60$\pm$0.26 &$-$0.95$\pm$0.09 &... &$>-$0.78$\pm$0.10$^a$\\
\hline 				       			 	 
\hline 				       			 	 
\end{tabular}			       			 	 
\end{center}			       			 	 
\begin{minipage}{140mm}
{\bf $^a$:} The abundance in Ca is a lower limit since the ionisation
potential of Ca II (11.868 eV) is lower than the one of HI and
therefore Ca II represents only some of the total calcium contained in a
neutral gas cloud.\\
\end{minipage}
\end{table*}

\subsection{Metal-Rich Systems}

The resulting abundances for each of the systems under study are
summarised in Table~\ref{t:ab}. All the absorbers are found to have
metallicities larger than the \nhi-weighted mean metallicity for
low-redshift DLAs. Overall the four quasar absorbers analysed here are
amongst the most metal-rich systems known to date (see also Prochaska
et al. 2006). The DLA with
\loghi=20.70 towards SDSS J0256$+$0110, in particular, has an almost
solar metallicity: [Zn/H]=$-$0.11$\pm$0.04. Furthermore, the sub-DLA
with \loghi=19.93 towards SDSS J0134$+$0051 studied here also yields
an upper limit [Zn/H]$<-$0.36 which is consistent with this absorber
being a metal-rich system. These results go in line with our recent
findings based on all sub-DLAs studied so far being better tracers of
the metallicity of the Universe because for a given dust-to-gas ratio,
there is less \hi\ gas and therefore dust in a sub-DLA than in a
DLA. Thus, the obscuration bias will affect the DLAs at a lower
dust-to-gas ratio as compared to the sub-DLAs (P\'eroux et al.,
2006). The net result is that the observed metallicity of sub-DLAs
will appear higher than those of classical DLAs in a given observed
limited magnitude quasar survey.

\subsection{Dust Content}

All the [Fe/H] measurements presented here are sub-solar
([Fe/H]$\sim$$-$1) whilst the [Zn/H] abundances determined are:
[Zn/H]=$-$0.11$\pm$0.04 in SDSS J0256$+$0110 at
\zabs=0.725, [Zn/H]=$-$0.54$\pm$0.20 in SDSS J1107$+$0048 at \zabs=0.740
and [Zn/H]=$-$0.49$\pm$0.22 in SDSS J2328$+$0022 at \zabs=0.652, plus one
upper limit ([Zn/H]$<-$0.36 in SDSS J0134$+$0051 at
\zabs=0.842). The ratio of refractory to volatile elements, such as 
 chromium and zinc respectively, provides an indication of the dust
content of quasar absorbers (e.g. Pettini et al. 1997). An increase in
the deficiency relative to the solar ratio [Cr/Zn]=0 corresponds to an
increase in the fraction of chromium depleted from the gas to the dust
phase. Here we derive [Cr/Zn]=$-$0.41 in SDSS J0256$+$0110,
[Cr/Zn]=$-$0.25 in SDSS J1107$+$0048 and [Cr/Zn]=$-$0.11 in SDSS
J2328$+$0022. So most of these systems are above the average estimate
of the dust in most quasar absorbers (see Meiring et al. 2006 for a
recent compilation), which is consistent with their being metal
rich. These results hint once more towards the existence of a
population of quasar absorbers having large amounts of dust and high
Zn abundance (Vladilo \& P\'eroux, 2005).

There have been suggestions of an obscured fraction of quasar
absorbers where the dustier systems would have dimmed their background
quasars to the extent than these would not be selected in abundance
studies in magnitude-limited samples. Boiss\'e et al. (1998)
empirically determined a so-called ``obscuration threshold'' of
[Zn/H]+$\log$\nhi=$\log$ N(Zn II)$>$13.15 in the [Zn/H] versus $\log$
\nhi\ parameter space above which no quasar absorbers were observed at
the time. But, there are now three DLA systems out of 65 known to lie
at $\log$ N(Zn II)$>$13.15, and of these three DLA/sub-DLA systems
only one deviates by more than 0.15 dex (SDSS J1323$-$0021; P\'eroux
et al. 2006). In the sample presented here, one additional system
(towards SDSS J0256$+$0110) is found to lie slightly above the
Boiss\'e dust cut-off. Therefore, the ``obscuration threshold''
reported by Boiss\'e et al. (1998) still appears to hold
empirically. It is interesting that the new points do not break the
envelope by much, given that other media are known to lie in this
region of the [Zn/H] versus $\log$ \nhi\ plot, such as Gamma Ray Burst
(GRB) hosts (see Savaglio et al. 2003, Watson et al. 2006). But it
might also be that GRB hosts, which are evidently warm, high density
regions, have strong Mg I and therefore overestimates of Zn II,
although a curve-of-growth analysis might on the contrary
underestimate Zn II.

Another mean of assessing the dust content of quasar absorbers is to
look at the possible reddening of the background quasar
continua. Here, we calculate the observed $g-i$ colours of the quasars
and derive the excess $\Delta (g-i)=(g-i)-(g-i)_{\rm med}$, where
$(g-i)_{\rm med}$ is the colour excess based on Sloan redshift
dependent quasar composite templates (which removes the effects of
redshift in the continuum and lines which affect the colors in a given
band). This leads to: $\Delta (g-i)$=$-$0.041, 0.081, 0.078 and
$-$0.137 for SDSS J0134$+$0051, SDSS J0256$+$0110, SDSS J1107$+$0048
and SDSS J2328$+$0022, respectively. Based on Figure 3 of York et
al. (2006), if $\Delta (g-i)$ $>$0.2, there is a good chance that the
object is reddened, while $\Delta (g-i)$ $<$0.2 is almost certainly
not reddened (because a random sample shows few objects at those
extremes). Therefore, there is no evidence for very strong reddening
due to the absorbers under study in any the four quasars, but the
$\Delta (g-i)$ values show that there is not inconsistency with taking
the object as reddened.

\subsection{D-index of Mg II Absorbers}

Because \nhi\ column density estimates at low-redshift require
space-based UV spectroscopy, there have been several attempts to find
accurate indicators of \nhi\ from metal lines parameters with large
rest-frame wavelengths (i.e. Rao, Turnshek \& Nestor 2006). More
recently, Ellison (2006) has proposed a finer indicator of the
presence of DLAs via the so-called $D$-index, i.e. the ratio of the
(Mg II $\lambda$ 2796) line's EW to velocity spread. Her result shows
that at fairly high-resolution \nhi\ and $D$-index correlate, with
$D$-index$>$6.3 corresponding to 'classical' DLAs. Our present sample
adds new data to further constrain the reported correlation: Mg II
$\lambda$ 2796 is covered for three systems (it falls in a spectral
gap for SDSS J2328$+$0022) plus an additional sub-DLA studied in
P\'eroux et al. (2006). We have measured the following $D$-indices for
SDSS J0134$+$0051, SDSS J0256$+$0110, SDSS J2328$+$0022 and SDSS
J1323$-$0021 respectively: $D$=5.6, 6.9, 8.5 and 7.8. These values
fall in the right ball-park of the $D$-index versus
\nhi\ plot of Ellison (2006) for the SDSS J0134$+$0051 sub-DLA and for
both the DLAs but is off in SDSS J1323$-$0021 case which admittedly is
a sub-DLA with a large \nhi\ column density. Therefore our data
confirm that the D-index is a good indicator of the presence of DLAs
among Mg II absorbers.

\subsection{Ca II Abundance at High-Resolution}

Along the same line of reasoning, there has been recent renewed
interest in the detection of Ca II H and K lines because they might
constitute a secure tool with which to detect DLAs at low-redshift
from medium-resolution optical spectra such as the ones from the Sloan
Survey (Wild \& Hewett, 2005; Wild, Hewett \& Pettini, 2006).  These
authors report a clear correlation between the strength of the Ca II
equivalent width and the amount of reddening in the background quasar
spectra.  These findings are surprising given that Ca II is known to
be highly depleted onto dust grains and that, therefore, one would
expect dusty absorber to have small {\it measured} Ca II
abundances. One way to better understand the relation between observed
Ca II and dust would to be study higher-redshift quasar absorbers
where the dust content can be estimated with other means (i.e. Cr
II/Zn II). To our knowledge, there are very few measurements of Ca II
in quasar absorbers with known N(HI) at high-redshift.  In our sample
of high-resolution spectra of \zabs$<$1 quasar absorbers (P\'eroux et
al., 2006 and the present paper), only SDSS J2328$+$0022 has the right
wavelength coverage to allow Ca II H and K detections. Both $\lambda$
3933 and $\lambda$ 3969 lines are nicely fitted at \zabs=0.652 toward
this quasar (see Figure~\ref{f:Q2328_others}) and we derive log N(Ca
II)=12.43$\pm$0.03 ([Ca/H]$>-$0.78$\pm$0.10).  Furthermore, we derive
[Ca/Zn]$>-$0.29.  More systems with a simultaneous measurement of Zn
II, Cr II and Ca II lines would help to understand the origin of the
trend between reddening and Ca II equivalent width.

\subsection{Possible Implications for Metallicity Evolution}

We now briefly discuss the implications of our data for the
metallicity evolution of DLAs. Given that the selection of these
systems is based on the presence of strong metal lines, these results
represent a ceiling to the global metallicity evolution.  This section
is based on DLAs only and therefore does not include the sub-DLA
towards SDSS J0134$+$0051 presented here or the one towards SDSS
J1323$-$0021 (P\'eroux et al., 2006). The analysis is based on the
statistical procedures outlined in Kulkarni \& Fall (2002), and uses
the data from this paper, the data compiled in Kulkarni et al. (2005),
our recent MMT data (Meiring et al. 2006), and other recent data from
the literature (Rao et al. 2005; Akerman et al. 2005). We binned the
combined sample of 111 DLAs in 6 redshift bins with 18 or 19 systems
each and calculated the global N$_{\rm H I}$-weighted metallicity in
each bin. Within each bin, we used Kaplan-Meier survival analysis to
account for the presence of some upper limits on Zn II column
densities. The $N_{\rm H I}$-weighted mean metallicity in the lowest
redshift bin $0.1 < z < 1.2$ is $- 0.79 \pm 0.15$. The linear
regression slope of the metallicity-redshift relation for the redshift
range $0.1 < z < 3.9$ is $-0.25 \pm 0.07$. The corresponding estimate
for the intercept of the metallicity-redshift relation is $-0.60 \pm
0.16$.

\begin{figure}
\begin{center}
\includegraphics[height=8cm, width=8cm, angle=0]{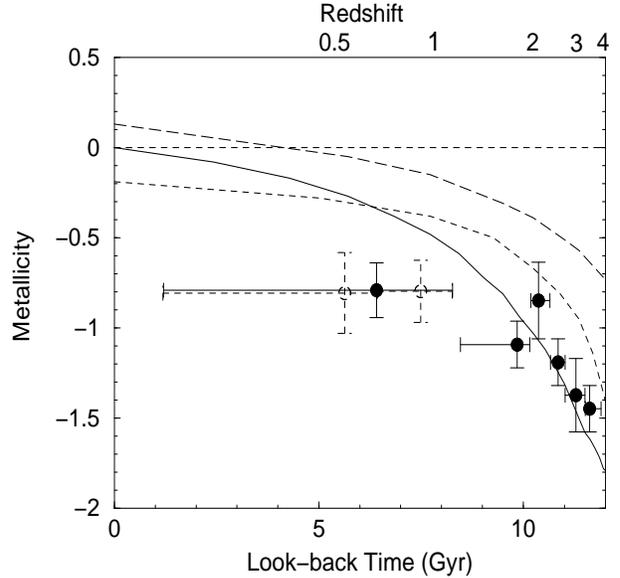}
\caption{$N({\rm H \, I})$-weighted mean Zn abundance relative to solar 
abundance vs. look-back time relation for 111 DLAs from our MMT,
Magellan, VLT, and HST data and the literature. 
Filled circles show 6
bins with 18 or 19 DLAs each.  Unfilled circles refer to the lowest
time bin split into 2 bins with 9 or 10 DLAs each. 
Horizontal bars denote ranges in look-back times covered by each
bin. Vertical error bars denote 1 $\sigma$ uncertainties including
sampling and measurement errors.  The solid, short-dashed and
long-dashed curves show, respectively, the mean metallicity in the
models of Pei et al. (1999), and Malaney \& Chaboyer (1996), and
Somerville et al. (2001). More metal-rich DLAs such as the ones
reported here would be required to make the global mean metallicity of
DLAs substantially deviate from sub-solar at low redshifts. For now,
the time evolution of the metallicity of DLAs seem to evolve slowly at
best. \label{f:Zlowz}}
\end{center}
\end{figure}

Figure~\ref{f:Zlowz} shows these results and compares them with
various models: Pei et al. (1999), Malaney \& Chaboyer (1996), and
Somerville et al. (2001) shown as solid, short-dashed and long-dashed
curves respectively. Pei et al. (1999) used a set of equations that
link the comoving densities of stars, interstellar gas, heavy
elements, and dust, averaged over the whole population of galaxies to
reproduce the properties of DLAs. Malaney \& Chaboyer (1996) carried
out chemical evolution calculations using the star formation history
from Pei \& Fall (1995) but without using the instantaneous recycling
approximation. Somerville et al. (2001) used semi-analytic models of
galaxy formation set within the cold dark matter merging hierarchy to
predict the metal content of the Universe. Based on empirical
relations, others (Chen et al. 2005; Zwaan et al. 2005) have argued
that a combination of metallicity-luminosity relation and metallicity
gradients in galaxy discs could produce lower mean metallicity at low
redshift, but most models predict that the abundance of DLAs at z=0
should be close to solar.

For now, the current samples support the conclusions of Khare et
al. (2004) and Kulkarni et al. (2005) that the metallicity of the
Universe traced by DLAs evolves slowly at z$<$3. This is clearly seen
in Figure~\ref{f:Zlowz} which shows the $N_{\rm H I}$-weighted mean
metallicity {\it vs.}  look-back time for each of our 6 redshift
bins. The fact that our new systems may be biased (due to the
selection criterion) towards high metallicity makes the above
observations even stronger. We also calculated the $N_{\rm H
I}$-weighted mean metallicity in two halves of the lowest time bin
since that bin encompasses a large time interval. The unfilled circles
in Figure~\ref{f:Zlowz} refer to the lowest time bin split into 2 bins
with 10 and 9 DLAs each. The points derived for these two half-bins
agree very closely with the one for the corresponding combined bin. In
all cases, there is a hint that the rate of metallicity evolution of
DLAs may have slowed down at $z < 2$, which might be interpreted as an
effect of dust obscuration. Figure~\ref{f:Zlowz} plotted versus age of
the Universe emphasizes once more the paucity of measurements of
abundances in quasar absorbers at the lowest redshifts, i.e. over the
last 8 Gyr of the Universe.
 
\section{Conclusions} 

To conclude, we have presented new high-resolution observations of Mg
I, Mg II, Fe II, Zn II, Cr II, Mn II, Ti II and Ca II towards 3 DLAs
and 1 sub-DLA at 0.6$<$\zabs$<$0.9 selected from the Sloan Digital Sky
Survey. While this small sample is not necessarily representative of
the general population given the pre-selection of strong metal lines,
it shows that there are high metallicity systems at z$<$1. We now know
of systems four times solar (P\'eroux et al. 2006) and between a half
and a third solar (this work), i.e. higher than the current z=0 mean
intercept of 1/6$^{th}$ solar. In addition, our measurements more than
double the sample of high-resolution Zn determinations at z$<$1 thanks
to the suitability of the blue-sensitive UVES spectrograph to these
type of studies.

A comparison of high-resolution spectra with lower resolution data in
a few cases where both data exist, has shown that problems of
particular blends at $\lambda
\lambda$ 2026 2062, and the possibility of saturation among so
many components, do not seem to preclude getting accurate values for
$N_{\rm ZnII}$ at lower resolution. Nevertheless, some Zn values in
the literature may be in error because of not having properly
addressed the potential contamination of Zn II $\lambda 2026$ with Mg
I near the same wavelength. In addition, using the better-determined
components in two of our systems, we find that [Fe/Zn] varies by a
factor of 5 and 6 respectively which is much more than what is
observed in DLAs at higher redshifts.

Based on [Cr/Zn] ratios and $\Delta (g-i)$ quasar continua reddening
estimates, two of the four systems are found to be probably fairly
dusty but interestingly only one of them marginally overcomes the
Boiss\'e et al. (1998) ``obscuration threshold''. We provide further
data which, for most of them, indicate that the $D$-index (EW of Mg II
compared to velocity width) is a sharper prediction of \nhi\ than EW(Mg
II) only. We measure the $D$-indices for four of our systems and find
most of them to lie on the \nhi\ versus $D$-index correlation plot
reported by Ellison (2006). In addition, we report the detection of
high-resolution Ca II lines towards SDSS J2328$+$0022 and discuss the
abundance measurement in the framework of the recent results of Wild,
Hewett \& Pettini (2006).

Finally, we show that relatively metal-rich systems exist in large
enough samples (see also Herbert-Fort et al. 2006), but that their
impact on the overall global metallicity will only be determined with
larger data samples. In particular, the possible impact of sub-DLAs on
these measurements will be reviewed in a forthcoming paper (Khare et
al., 2006). It is a possibility that these systems are less prone to
the biasing effect of dust and thus represent the only tool currently
available to detect the most metal-rich galaxies seen in
absorption. More such measurements at z$<$1 are required to be able to
accurately assess the role played by these metal-rich systems in the
global evolution of metals.

\section*{Acknowledgments}

We would like to thank the Paranal and Garching staff at ESO for
performing the observations in Service Mode. JDM and VPK acknowledge
partial support from the U.S. National Science Foundation grant
AST-0206197 (PI: Kulkarni).

\end{document}